# douka: A universal platform of data assimilation for materials modeling

Aoi Watanabe[1]*, Ryuhei Sato[1], Ikuya Kinefuchi[2], Yasushi Shibuta[1]

[1]Department of Materials Engineering, The University of Tokyo,

7-3-1 Hongo, Bunkyo-ku, Tokyo 113-8656, Japan

[2]Department of Mechanical Engineering, The University of Tokyo,

7-3-1 Hongo, Bunkyo-ku, Tokyo 113-8656, Japan

**Abstract**

A large-scale, general-purpose data assimilation (DA) platform for materials modeling, douka, was developed and applied to nonlinear materials models. The platform demonstrated its effectiveness in estimating physical properties that cannot be directly obtained from observed data. DA was successfully performed using experimental images of oxygen evolution reaction at a water electrolysis electrode, enabling the estimation of oxygen gas injection velocity and bubble contact angle. Furthermore, large-scale ensemble DA was conducted on the supercomputer Fugaku, achieving state estimation with up to 8,192 ensemble members. The results confirmed that runtime scaling for the prediction step follows the weak scaling law, ensuring computational efficiency even with increased ensemble sizes. These findings highlight the potential of douka as a new approach for data-driven materials science, integrating experimental data with numerical simulation.

*Corresponding Author. Tel.: +81-3-5841-7119, E-mail: watanabe@mse.mm.t.u-tokyo.ac.jp (A. Watanabe)

## 1. Introduction

As a new paradigm in materials science, the data-driven approach, which aims to discover new insights from experimental and simulation data, is gaining attention [1]. Machine learning (ML) methods are a typical example of such methods. The integration of ML techniques with the development of diverse materials databases has successfully revealed latent correlations and characteristics within materials science data [2–4]. However, ML-based regression and classification often obscure the underlying physics of the data. Consequently, significant efforts have been devoted to developing and applying ML methods that incorporate physical principles [5]. In contrast, many phenomena in materials science can be explained through governing equations that encapsulate their physical principles (i.e., physics-based models). For instance, heat and mass diffusion can be described by the heat conduction equation (diffusion equation), while fluid dynamics can be modeled using the Navier-Stokes equations. Integrating physics-based models with experimental data can facilitate a clearer understanding of underlying physical principles and theories while leveraging established models. This approach offers a valuable and distinctive alternative to ML, which often derives relationships in a black-box manner.

In recent years, data assimilation (DA) [6] has emerged as a promising approach. DA is a statistical estimation technique that integrates numerical models and simulations with experimental data and was initially developed in meteorology and oceanography [7]. Today, it is widely applied in various practical fields, including weather and ocean forecasting [8–10]. More recently, its applications have expanded to materials science, encompassing various phenomena such as microstructure formation in metallic materials [11,12], solid-phase sintering [13], crystal growth of two-dimensional materials [14], crystal structure determination [15], and the prediction of material properties [16–18].

The DA method is fundamentally a framework that operates independently of the specific physical models and governing equations employed. However, most previous DA studies in materials science have incorporated DA methods tailored to the governing equation of interest (hereafter referred to as the material model) for each specific study. Furthermore, understanding DA requires proficiency in probability and statistics, which can be challenging for many materials science researchers to grasp and apply. These factors have impeded the widespread adoption of data assimilation in materials research. Given this background, we propose that standardizing components other than the material model would facilitate the broader application of DA in materials science. To this end, this research aims to develop a versatile and scalable DA platform applicable to various material models, contributing to the advancement of data-driven science for understanding complex physical phenomena in materials research.

## 2. Outline of Data Assimilation

### 2.1 Sequential and non-sequential data assimilation

DA methods are techniques used to incorporate observational data into governing equations based on physical phenomena to perform state estimation. These methods are broadly classified into two types: sequential and non-sequential DA. Sequential DA refers to a method in which observational data is provided as a time series and is sequentially incorporated into numerical simulations. Representative sequential DA methods include the Kalman Filter (KF) [19], the ensemble Kalman Filter (EnKF) [20], and the particle filter [21]. Sequential DA is particularly suitable for predicting states or estimating time-varying states. Non-sequential DA, on the other hand, refers to a method used when observational data over a specified time period is provided in advance. This method estimates an initial state that best represents the system by integrating the given observational data. Representative non-sequential DA methods include the three-dimensional variational method (3DVar) [22] and the four-dimensional variational method (4DVar) [23]. In this study, we focus on sequential DA because many experiments in the field of materials science involve time-series observational data such as the formation process of materials and the dynamic behavior of microstructures. Since most material models consist of nonlinear equations, this study employs the EnKF, which can accommodate nonlinear models. The details of the EnKF are described later.

**2.2 Procedure of sequential data assimilation**

In a conventional numerical simulation, the state vector $\bm{x}_t$, which consists of physical variables at time *t*, takes a unique value, and its sequential changes are deterministic. This is called a simulation model and can be expressed as follows:

$$\bm{x}_t = f_t(\bm{x}_{t-1}) \tag{1}$$

In DA, on the other hand, a distribution of errors is introduced into the simulation results, and the state vector $\bm{x}_t$ is represented by a probability density function. This is called a system model.

$$\bm{x}_t = M_t(\bm{x}_{t-1}, \bm{v}_t) \tag{2}$$

$\bm{v}_t$ is called the system noise at time *t*, representing the uncertainty of the prediction due to imperfections in the numerical simulation model, errors caused by numerical calculations, and other factors. Furthermore, considering that the observed data also contain errors, the observation model with observation noise $w_t$ is represented by a probability density function.

$$y_t = h_t(x_t) + w_t \tag{3}$$

where $y_t$ is the observation vector and $h_t$ is the observation operator, which transforms the state vector $x_t$ into a form comparable to the observed values. The system model (Eq. (2)) and the observation model (Eq. (3)) together are referred to as the state-space model.

Sequential DA consists of two steps: the prediction step and the filtering step. In the prediction step, the state one step ahead, $x_{t|t-1}$, is estimated from the state $x_{t-1|t-1}$ based on the system model as given in Eq. (2). The subscript notation $x_{i|j}$ represents the state at time step *i* in the numerical simulation, incorporating observed data up to observation time *j*. Next, in the filtering step, the observational data at time *t*, $y_t$ is used to modify the estimated state based on Eq. (3), yielding the updated state $x_{t|t}$. Therefore, by taking *i* and *j* as the *x*- and *y*-axes, respectively, the process of sequential DA can be represented as shown in Figure 1. By repeating the two steps of prediction and filtering, the estimation of the state is achieved by sequentially incorporating the observed data in the sequential DA. For further details on the principle of DA, see Ref. [6].

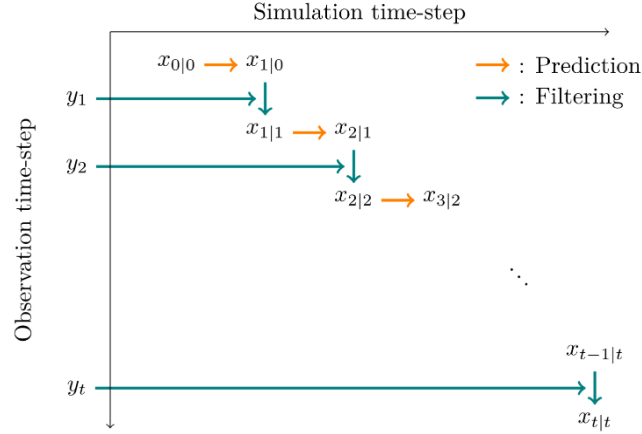

**Fig. 1.** Schematic diagram of the process of sequential data assimilation. Horizontal (*x*) and vertical (*y*) axes represent time-step of simulation and observation, respectively.

### 2.3 Kalman filter (KF)

When both the system model and the observation model are linear, and all probabilistic variables, including system noise and observation noise, follow a normal distribution, the Kalman filter [19] described below can be applied. In this case, the state-space model is expressed as follows

$$\boldsymbol{x}_t = \boldsymbol{M}_t \boldsymbol{x}_{t-1} + \boldsymbol{v}_t \tag{4}$$

$$\boldsymbol{y}_t = \boldsymbol{H}_t \boldsymbol{x}_t + \boldsymbol{w}_t \tag{5}$$

$\boldsymbol{M}_t$ and $\boldsymbol{H}_t$ represent the simulation model and the observation operator, respectively, and are expressed as matrices in the Kalman filter since both are linear. Consider a state-space model, in which the state vector is *k*-dimensional and the observation vector is *l*-dimensional. The vector of state mean value $\boldsymbol{x}_{t|t}$ and the observation vector $\boldsymbol{y}_t$ are defined as follows:

$$\boldsymbol{x}_{t|t} = \left\{ x_{t|t}^{(0)} \quad x_{t|t}^{(1)} \quad \cdots \quad x_{t|t}^{(k-1)} \right\}^T \tag{6}$$

$$\boldsymbol{y}_t = \left\{ y_t^{(0)} \quad y_t^{(1)} \quad \cdots \quad y_t^{(l-1)} \right\}^T \tag{7}$$

Let $Q_t$ and $R_t$ be the variance-covariance matrices of the system noise $v_t$ and the observed noise $w_t$, respectively. In the prediction step of the Kalman filter, the vector of state mean value $x_{t|t-1}$ and its covariance matrix $V_{t|t-1}$ is defined as:

$$x_{t|t-1} = M_t x_{t-1|t-1} \tag{8}$$

$$V_{t|t-1} = M_t V_{t-1|t-1} M_t^T + Q_t \tag{9}$$

Then, the filtering step is defined as:

$$x_{t|t} = x_{t|t-1} + K_t(y_t - H_t x_{t|t-1}) \tag{10}$$

$$V_{t|t} = (I - K_t H_t) V_{t|t-1} \tag{11}$$

where $K_t$ is referred to as the Kalman gain and is defined as follows:

$$K_t = V_{t|t-1} H_t^T \left( H_t V_{t|t-1} H_t^T + R_t \right)^{-1} \tag{12}$$

The covariance matrix of the system noise $Q_t$, is calculated as follows

$$Q_t = \begin{pmatrix} \sigma_0^2 & \sigma_{0,1} & \cdots & \sigma_{0,k-1} \\ \sigma_{1,0} & \sigma_1^2 & & \sigma_{1,k-1} \\ \vdots & & \ddots & \vdots \\ \sigma_{k-1,0} & \sigma_{k-1,1} & \cdots & \sigma_{k-1}^2 \end{pmatrix} \tag{13}$$

The diagonal elements $\sigma_i^2$ (for $i \in [0, k-1]$) represent the variances of the system noise and the off-diagonal elements $\sigma_{i,j}$ (for $i, j \in [0, k-1]$ and $i \neq j$) represent the covariances between the system noise. The covariance matrix of the observed noise $R_t$ is defined similarly. Since all these processes involve linear matrix transformations, the time evolution of the state vector can be determined analytically, which is a key advantage of the Kalman filter.

## 2.4 Ensemble Kalman filter (EnKF)

When the governing equations are nonlinear, the limitation of the Kalman filter is that the time evolution of the state vector's variance, induced by the system model, cannot be directly calculated during the prediction process. To address this limitation, methods such as ensemble approximations are often employed. This approach is generally referred to as the ensemble Kalman Filter (EnKF) [20]. EnKF approximates the distribution of $x_t$ using a set of $N$ samples as follows

$$p(x_t) \simeq \frac{1}{N} \sum_{i=1}^{N} \delta(x_t - x_t^{(i)}) \qquad (14)$$

where $\delta$ is the delta function. The set of samples $\{x_t^{(i)}\}_{i=1}^{N}$ is called the ensemble and $x_t^{(i)}$ is referred to as the ensemble member. Specifically, when the number of ensemble members is $N$, the vector of state mean value of a sample $x_{t|t}$, which is $k$ dimensional, is expanded into an $k \times N$ dimensional matrix $X_{t|t}$ consisting of $N$ samples.

$$X_{t|t} = \left\{ x_{t|t}^{(0)} \quad x_{t|t}^{(1)} \quad \cdots \quad x_{t|t}^{(N-1)} \right\} \qquad (15)$$

For the prediction step, the system model expressed in Eq. (2) is applied to each ensemble member individually, which is equivalent to running $N$ simulations simultaneously. The state vector variance $V_{t|t-1}$ based on ensemble approximation can be computed as follows:

$$\breve{x}_{t|t-1}^{(i)} = x_{t|t-1}^{(i)} - \frac{1}{N} \sum_{j=0}^{N-1} x_{t|t-1}^{(j)} \qquad (16)$$

$$\breve{X}_{t|t-1} = \left\{ \breve{x}_{t|t-1}^{(0)} \quad \breve{x}_{t|t-1}^{(1)} \quad \cdots \quad \breve{x}_{t|t-1}^{(N-1)} \right\} \qquad (17)$$

$$V_{t|t-1} = \frac{1}{N-1} \breve{X}_{t|t} \breve{X}_{t|t}^T \qquad (18)$$

For the filtering step, the same procedure as the Kalman filter is applied to obtain Kalman Gain $K_t$ (Eq. (12)). From this, the filtered ensemble member $x_{t|t}^{(i)}$ is computed as follows:

$$x_{t|t}^{(i)} = x_{t|t-1}^{(i)} + K_t\left(y_t + \breve{w}_t^{(i)} - H_t x_{t|t-1}^{(i)}\right) \quad (19)$$

Note that, the observation noise factor $\breve{w}_t^{(i)}$ is added to the filtered ensemble member in EnKF. Here, $\breve{w}_t^{(i)}$ is computed from the observation noise $w_t^{(i)}$ as follows:

$$\breve{w}_t^{(i)} = w_t^{(i)} - \frac{1}{N}\sum_{j=0}^{N-1} w_t^{(j)} \quad (20)$$

## 3. Construction of douka platform

### 3.1. Structure

As described above, DA can be applied to various nonlinear material models using an ensemble approximation. Therefore, having a general-purpose platform that facilitates the easy integration of various existing material models would be highly beneficial. Moreover, ensemble approximation requires a sufficiently large number of ensemble members, *N*, to accurately capture the time-dependent variation of variance. Therefore, we construct douka, a large-scale, general-purpose DA platform in this study, with a focus on these two key aspects. To enhance versatility and performance, this system implements the prediction and filtering steps as separate individual commands. Figure 2 represents schematic representation of the structure for parallel data assimilation execution flow using individual process commands in douka. First, for the prediction step, the subcommand `predict` is implemented, allowing the material model to be introduced as a plugin. This enables any existing material simulation model to be easily incorporated into the DA framework. Next, the subcommand `filter` is implemented to perform filtering using the observation vector on the state vector estimated during the prediction step. Since the filtering process is model-independent, the system ensures platform versatility by structuring DA processes separately. This means that users can apply their own simulation models without reimplementing the filtering process.

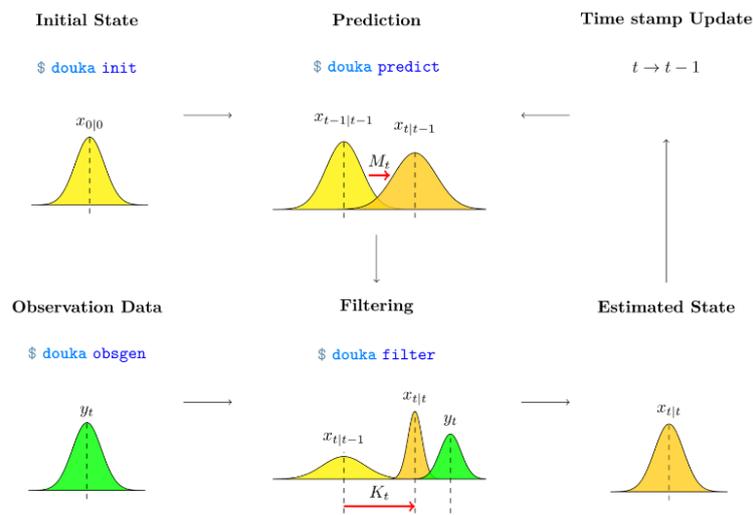

**Fig. 2.** Schematic representation of the parallel data assimilation execution flow using individual process commands in douka.

### 3.2. Parallelization of ensemble

To improve the performance of the DA platform, implementing ensemble parallelization is crucial. Therefore, the `predict` command is designed as a process that executes the prediction step for each ensemble. These processes run in parallel to achieve ensemble parallelization. At this stage, the system transfers state vectors via the file system. Figure 3(a) illustrates a schematic diagram of the system architecture of a typical supercomputer. As shown, a shared file system is constructed within the local network of the supercomputer. By transferring state vectors through this shared file system, the platform achieves ensemble parallelization even on a supercomputer. Figure 3(b) depicts the sequence of DA, including the prediction and filtering steps, via the shared file system, proposed in douka. The process flow is as follows: Each node executes the prediction step for each ensemble and writes the results to files. Then, the filtering process reads the predicted ensemble $X_{t|t-1}$ and the given observation vector $y_t$ to compute the posterior ensemble $X_{t|t}$. As a result, the entire DA computation is divided into a collection of smaller jobs. Additionally, this method enables the file output of the state vector to serve as a checkpoint in the DA sequence.

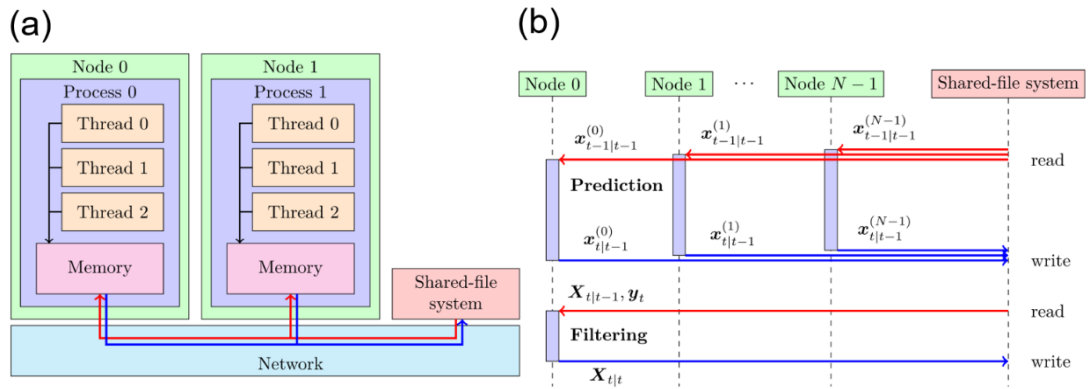

**Fig. 3.** (a) Schematic diagram of the system architecture of a typical supercomputer. (b) Sequence diagram of the data assimilation process in a supercomputer.

### 3.3. Procedure for executing each process in douka

In douka, following commands allow DA procedures to be executed as individual processes.

```
$ douka --help
douka [Command]

Command:
   init         Provide initial distribution
   predict      Prediction step for an ensemble model
   filter       Filter state vectors with observation data
   obsgen       Generate observation data for twin experiment
```

```
Options:
   --help      (Opt) Print help message
   --version   (Opt) Print version
```

The `init` command generates the initial distribution for each ensemble, and the `obsgen` command yields the observation data for the twin experiment. The prediction step (`predict` command) and filtering step (`filter` command) are explained in the following sections.

### 3.3.1 File Naming rule

The state vectors handled in douka follow the JSON format [24], and their file names follow the naming rule described below:

| | |
|---|---|
| **State Vector File**: | ${NAME}_${ID}_${SYS_TIM}_${OBS_TIM}.json |
| **Observation Vector File**: | ${NAME}_obs_${OBS_TIM}.json |

The variables used are explained as follows:

**NAME**: An arbitrary name specified by the user (e.g., `lorenz63`).
**ID**: Ensemble ID (e.g., `0001`). It is expressed with four digits, padded with zeros. If the total number of ensemble members is $N$, the values range from $0$ to $N-1$.
**SYS_TIM**: The system time (e.g., `000000`). It is expressed with six digits, padded with zeros.
**OBS_TIM**: The observation time (e.g., `000001`). It is expressed with six digits, padded with zeros.

The contents of the state vector file and the observation vector file are shown in Sec. 3.1 and 3.2 in User Guide provided as Supplementary Information.

### 3.3.2 Predict command

The execution of prediction step is computed by the following equation:

$$x_{t|t-1}^{(i)} = M_t\left(x_{t-1|t-1}^{(i)}, v_t\right) \tag{21}$$

Here, the system noise $v_t$ is computed using the covariance matrix $Q_t$, provided by the user, and the seed value $s_{\text{predict}}$ used for random number generation. The input and output of the `predict` command are shown in Table 1. The material model $M_t$ is dynamically loaded as a shared library at runtime. Additionally, using the same seed value for each ensemble $i$ and each simulation time $t_{\text{sim}}$ could introduce statistical bias. To prevent this, the actual seed value $s_{\text{predict}}$ used is derived by the following equation, using the input seed value $s_{\text{in}}$, simulation time $t_{\text{sim}}$, and ensemble ID $i$:

$$s_{\text{predict}} = s_{\text{in}} + t_{\text{sim}} + i \tag{22}$$

These parameters are provided through a JSON file and input via the following command to execute the prediction process. The details of the parameters used here are shown in Sec. 6.1 in User Guide provided as Supplementary Information.

```
$ douka predict --help
douka predict [Options]

Options:
   --state          Input state vector json file
   --param          Input parameter json files
   --plugin         System model plugin
   --plugin param   (Opt) Plugin option json file
   --output         (Opt) Output path (default='output')
   --force          (Opt) Overwrite existing file
   --help           (Opt) Print help message
```

In this command, the `--plugin` option specifies the dynamic library of the material model to be executed during the prediction process. The implementation method for the plugin's dynamic library is described in Sec. 3.4. By implementing a plugin, various material models can be easily integrated into the DA platform. In addition, the parameters used in the plugin are specified by the `--plugin_param` option.

**Table 1.** Input and output for prediction process

|  | Variable | Description |
|---|---|---|
| Input | $x^{(i)}_{t-1|t-1}$ | Ensemble member for prediction |
|  | $M_t$ | Material model plugin |
|  | $Q_t$ | System noise covariance matrix |
|  | $k$ | Dimension of the state vector |
|  | $s_{in}$ | Seed value |
| Output | $x^{(i)}_{t|t-1}$ | Predicted ensemble member |

### 3.3.3 Filter command

Filtering step is calculated by the Eq. (19) based on EnKF scheme. The Kalman gain, $K_t$ is the same as Eq. (12). The observation noise, $w_t$, is calculated using the covariance matrix $R_t$ and the seed value used for random number generation, both of which are given by the user. Also, the variance of the state vector, $V_{t|t-1}$, is computed from $X_{t|t-1}$ using Eq. (18). The inputs and outputs of `filtering` command are as shown in Table 2. The observation vector, $y_t$, is assumed to be prepared in advance through experiments. As the input ensemble, the outputs of the aforementioned prediction step are used. Here, while the each ensemble member $x^{(i)}_{t-1|t-1}$ was calculated in the prediction step, the entire ensemble $X_{t|t}$ is handled collectively in the filtering step. In addition, similar to the prediction step, the seed value $s_{obs}$ used for random number generation is as follows, using the observation time $t_{obs}$:

$$s_{\text{obs}} = s_{\text{in}} + t_{\text{obs}} + i \tag{23}$$

The details of the parameter used for filtering is shown in Sec. 7.1 in User Guide provided as Supplementary Information.

```
$ douka filter --help
douka filter [Options]

Options:
   --state       Input state vector json file
   --param       Input parameter json files
   --obs         Input observation json file
   --filter      (Opt) Filter [enkf|particle] (default=enkf)
   --output      (Opt) Output path (default='output')
   --force       (Opt) Overwrite existing file
   --help        (Opt) Print help message
```

**Table 2.** Input and output for filtering process

|        | Variable      | Description                         |
|--------|---------------|-------------------------------------|
| Input  | $X_{t|t-1}$   | Ensemble for filtering              |
|        | $y_t$         | Observation vector                  |
|        | $R_t$         | Observation noise covariance matrix |
|        | $H_t$         | Observation matrix                  |
|        | $k$           | Dimension of the state vector       |
|        | $l$           | Dimension of the observation vector |
|        | $s_{\text{in}}$ | Seed value                        |
| Output | $X_{t|t}$     | Filtered ensemble                   |

### 3.4. Implementation of plug-ins

The implementation of the plugin for this platform is described in the programming language C++. Initially, the `douka::PluginInterface` class is inherited, and implement the following two methods. The first method is the `set_option`, which sets parameters to be used in the system model as needed. The second method is the `predict`, which performs a prediction step based on desired simulation model. The first input argument of this method is a reference to a vector of state mean value, which is updated by the material model as state prediction. An example plug-in implementation of the Lorenz63 model (Section 4.1) is shown in Sec. 2 in User Guide provided as Supplementary Information. The plug-in implementation of other examples (for Section 4.2 and 4.3) can be obtained at GitHub. It is built as a dynamic library and used as a system model for DA by specifying it with the `--plugin` option of the `douka` command. A key contribution of the proposed DA platform lies in its model agnostic architecture. Specifically, the platform is designed such that sole model-specific customization necessary for integrating diverse material models is the implementation of a state vector update within the `predict` method.

### 3.5. Ways to execute the entire process of data assimilation

The DA process alternately executes the `predict` and `filter` commands to assimilate data while outputting the state vector to a file as show in Figure 4. The example code for Lorenz63 model is shown below. Here, the '&' at the end of the `douka predict` command allows the process to run in the background, enabling parallel execution of each process in the prediction ensemble. In addition, the process can be executed on a supercomputer by replacing the execution part of each process with a job script submission under the flow of this process. In this case, ensemble parallelization of prediction can be realized by bulk job execution. For further details on execution, see User Guide provided as Supplementary Information.

```
# Bash-like pseudo code for DA process
N = 3
T = 10

for (( t = 0; t < T; t++)); do
  # Set timestamp
  SYS_TIM=$(printf "%06d" $t)
  OBS_TIM=$(printf "%06d" $t)
```

```
  # Prediction
  for (( i = 0; i < N; i++)); do
    douka predict ¥
      --
state  ./output/lorenz63 ${i} ${SYS TIM} ${OBS TIM}.json ¥
      --param  lorenz63.predict.json ¥
      --plugin ./lorenz63.so &
  done
  wait

  # Filtering
  SYS TIM=$(printf "%06d" $(( t + 1 )))
  douka filter ¥
      --state  lorenz63 %04d ${SYS TIM} ${OBS TIM}.json ¥
      --param  lorenz63.filter-enkf.json ¥
      --obs    lorenz63 obs ${SYS TIM}.json ¥
Done
```

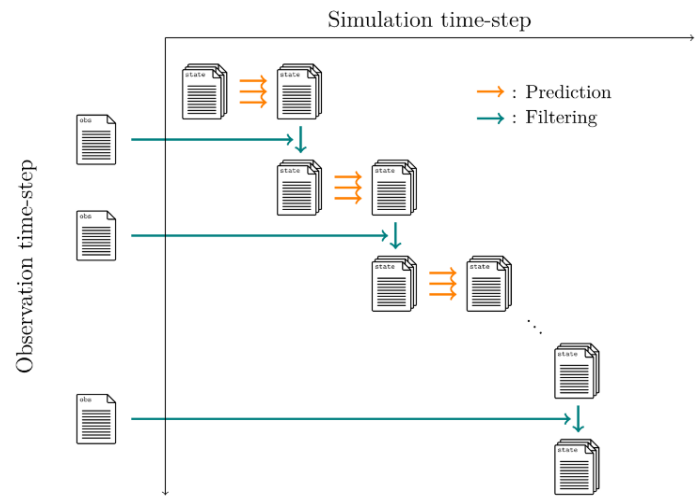

**Fig. 4.** Schematic diagram of file updates during data assimilation process execution.

### 3.6. Acceleration techniques

The DA processing, as depicted in Eqs. (12) and (19), involves intensive matrix operations. For these implementations, Eigen [25] library is selected due to its computational efficiency and flexibility. To explore performance improvements, established matrix calculation libraries, such as BLAS [26] and LAPACK [27], can be integrated as Eigen backends. Specifically, the Intel Math Kernel Library (MKL) [28], which provides thread-parallelized BLAS and LAPACK routines, can be utilized to enhance performance.

## 4. Example applications

### 4.1. Lorenz63 model for non-linear problem

In this section, we present a series of DA applications to various models to demonstrate the versatility of the douka platform developed in this study. We begin with the Lorenz63 model [29], a simple system that exhibits chaotic behavior. The Lorenz63 model is an idealized representation of thermal convection, described by the following three differential equations:

$$\frac{\partial x}{\partial t} = \sigma(y - x) \tag{24}$$

$$\frac{\partial y}{\partial t} = x(\rho - z) - y \tag{25}$$

$$\frac{\partial z}{\partial t} = xy - \beta z \tag{26}$$

where $x$, $y$ and $z$ represent the intensity of convection, the horizontal temperature difference, and the vertical temperature difference, respectively. The parameters $\sigma$, $\rho$, and $\beta$ are parameters related to the Prandtl number, Rayleigh number, and certain physical dimensions of the layer, respectively. Here, we set their values to $\sigma = 10$, $\rho = 32$, and $\beta = 2.665$. It is well known that the system exhibits chaotic behavior near these parameter values. Eqs. (24), (25), and (26) are discretized using the finite difference method with a time step of 0.01 second to numerically compute the system's time evolution. Figure 5 illustrates the trajectories of the Lorenz63 model over 1000 time steps for two sets of initial condition: $(x_0, y_0, z_0) = (1.0, 3.0, 5.0)$ and $(x_0, y_0, z_0) = (1.1, 3.3, 5.5)$. A mere 10% difference in the initial conditions leads to a significant divergence in the trajectories after 1000 time steps, clearly demonstrating the model's chaotic nature.

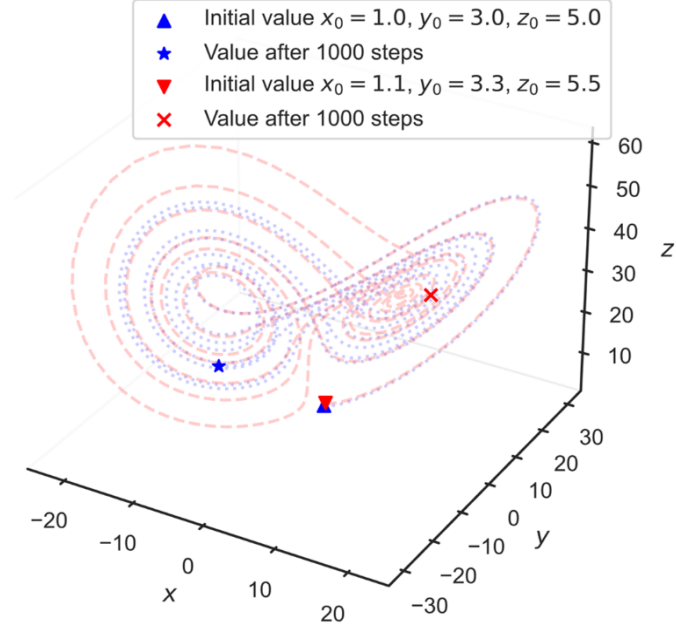

**Fig. 5.** Trajectories of the Lorenz63 model with different initial conditions of $(x_0, y_0, z_0) = (1.0, 3.0, 5.0)$(blue line) and $(x_0, y_0, z_0) = (1.1, 3.3, 5.5)$(red line).

Here, DA is performed using the results of numerical simulation of the Lorenz63 model with the initial state $(x_0, y_0, z_0) = (1.0, 3.0, 5.0)$ as the observed data. This type of DA in which data generated by numerical simulations are regarded as synthetic observed data is called a twin experiment. Twin experiments are commonly used to evaluate the applicability of DA algorithms. Here, the initial values are assumed to be unknown and the state of the Lorenz63 model, which exhibits chaotic behavior, is estimated. The result of the numerical simulation from the initial values described above, plus the error due to the variance-covariance matrix $\boldsymbol{R}_0 = \mathrm{diag}(0.1, 0.1, 0.1)$, were used as the synthetic observed data. The state and observation vectors are defined in terms of the variables $x$, $y$ and $z$ from Eqs. (23) to (25) as follows:

$$\boldsymbol{x}_{t|t} = \{\hat{x} \quad \hat{y} \quad \hat{z}\}^T \tag{27}$$

$$\boldsymbol{y}_t = \{x \quad y \quad z\} \tag{28}$$

The circumflex symbols in Eq. (27) represent estimated values and the same notation applies hereafter. For prediction, $M_t(\boldsymbol{x}_{t|t})$ corresponds to computing the value after one time step by approximating Eqs. (23) to (25) using the forward differencing method, yielding the predicted state $\boldsymbol{x}_{t+1|t}$. The initial state distribution is assumed to be a multivariate normal distribution with mean value $(x_0, y_0, z_0) = (-1.0, -3.0, -5.0)$ and variance of 20.0 for each component. It should be noticed that these initial conditions for the estimation process are distinct from those used to generate the synthetic observational data, which yield a significant difference in the behavior of the Lorentz63 model solution. Parameters for the DA are shown in Table 3. The system noise covariance $\boldsymbol{Q}_t$ and the observation noise covariance $\boldsymbol{R}_t$ are constructed as follows using the values in Table 3.

$$\boldsymbol{Q}_t = \text{diag}(Q_x, Q_y, Q_z) \tag{29}$$

$$\boldsymbol{R}_t = \text{diag}(R_x, R_y, R_z) \tag{30}$$

All calculations in this study were performed on the workstation equipped with two Intel® Xeon® w7-3445 CPUs and 64 GiB memory, except for those conducted on the supercomputer Fugaku in Section 4.4.

**Table 3.** Data assimilation parameters for Lorenz63 model

| | | | |
|---|---|---|---|
| N | Number of ensemble members | | 100 |
| k | Dimension of the state vector | | 3 |
| l | Dimension of the observation vector | | 3 |
| $\Delta t_{obs}$ | Observation time interval | 0.1 | [sec] |
| $Q_w$ | System noise covaiance for $w \in (x, y, z)$ | 5.0 | |
| $R_w$ | Observation noise covariance for $w \in (x, y, z)$ | 10.0 | |
| $\boldsymbol{H}_t$ | Observation matrix | $\boldsymbol{I}_{3,3}$ | |

Figure 6 illustrates the differences in state estimation with and without DA in the Lorenz63 model. The crosses mark the observed data, and there are a total of 21 points at 0.1-second intervals up to 2 seconds. The upper panels ((a) to (c)) show the results without DA and the lower panels ((d) to (f)) show the results with DA, which are displayed separately for each x, y, and z direction for visibility. In the without DA case, the initial state for the simulation is largely different from that for the synthetic observed data, and thus the time evolution of the state is very different. On the other hand, with DA, the time evolution of the state follows the observed data in all 100 ensemble members because the filtering process makes appropriate corrections at the time the observed data are present. This twin experiment confirmed that douka can be used to achieve appropriate data DA for models with strong nonlinearities.

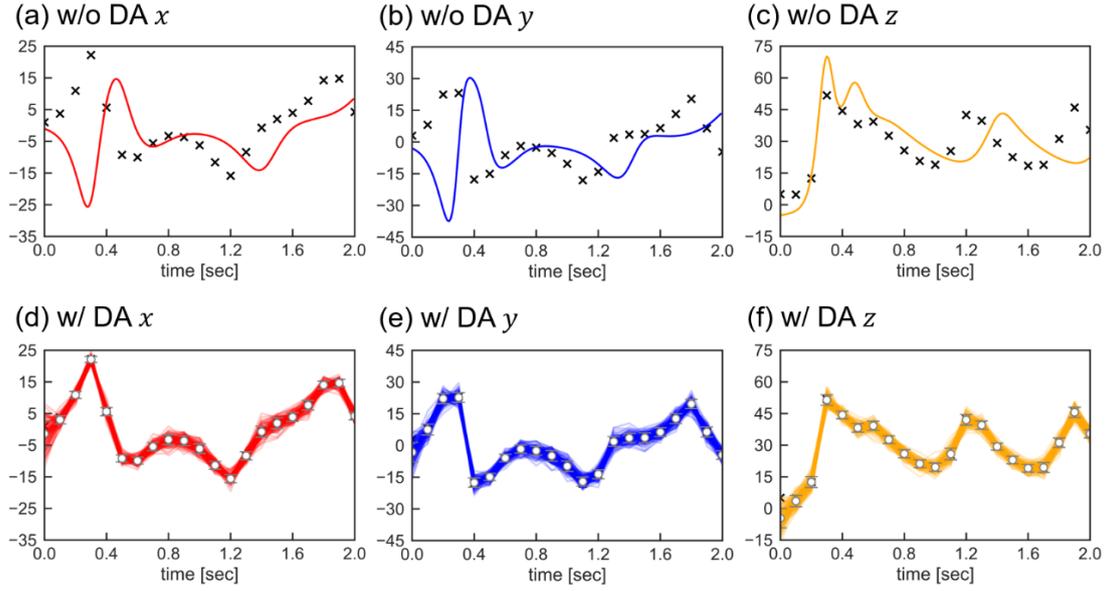

**Fig. 6.** Comparison of state estimation results in the Lorenz63 model: (a)–(c) without data assimilation (DA) and (d)–(f) with DA. Crosses indicate synthetic observation data, while white dots denote the estimated mean.

**4.2 Phase-field model for free-boundary problem**

The phase-field model (PFM) has been widely applied in the field of materials science [30], including the dendrite growth during solidification [31,32] and the grain growth [33]. Previous DA researches using phase-field models have contributed to the estimation of interfacial properties [16,34,35] and the improvement of model accuracy [13]. In this section, we employ the PFM to simulate the solidification of pure Ni and use DA for the estimation of anisotropy of interfacial parameters. In the PFM, the morphology of the microstructure is defined by the spatial distribution of the phase-field variable $\phi$. It is known that the numerical results of PFM are affected by the interface width, particularly when simulating system with temperature gradients across the interface. To address this issue, a quantitative PFM that accounts for temperature variations within the interface region are proposed [36]. We employ the quantitative PFM [36] to model the solidification process of pure Ni. In this framework, $\phi = 1$ and $\phi = -1$ correspond to the solid and liquid phases, respectively, with the solid–liquid interface characterized by a continuous transition between −1 and 1. The time evolution equations of $\phi$ are presented below.

$$\tau \frac{\partial \phi}{\partial t} = \nabla(W\nabla\phi) + \sum_{w=x,y} \frac{\partial}{\partial w}\left(|\nabla\phi|^2 W \frac{\partial W}{\partial\left(\frac{\partial \phi}{\partial w}\right)}\right) + \phi - \phi^3 - \lambda(\phi - \phi^2)u \quad (31)$$

The variables $\tau$ and $W$ are functions of the normal direction $\boldsymbol{n}$ of the phase field variable $\phi$, and are defined by the following expressions:

$$W(\boldsymbol{n}) = W_0 a_c \quad (32)$$

$$\tau(\boldsymbol{n}) = \tau_0 a_c a_k \quad (33)$$

$$a_c(\boldsymbol{n}) = (1 - \varepsilon_c)\left\{1 + \frac{4\varepsilon_c}{1 - 3\varepsilon_c}(n_x^4 + n_y^4)\right\} \quad (34)$$

$$a_k(\boldsymbol{n}) = (1 + \varepsilon_k)\left\{1 - \frac{4\varepsilon_k}{1 + 3\varepsilon_k}(n_x^4 + n_y^4)\right\} \quad (35)$$

$$n_x^4 + n_y^4 = \frac{\left(\frac{\partial \phi}{\partial x}\right)^4 + \left(\frac{\partial \phi}{\partial y}\right)^4}{|\nabla\phi|^4} \quad (36)$$

To represent crystallographic anisotropy of Ni, which exhibits a 4-fold symmetric crystal structure, the term $n_x^4 + n_y^4$ is employed. Here, $\varepsilon_c$ and $\varepsilon_k$ correspond to the interface anisotropy and kinetic anisotropy, respectively, and are crucial parameters influencing the morphology during solidification. Figure 7 depicts the simulation result of quantitative PFM for Ni solidification from initial solid seed using parameters listed in Table 4. Since the anisotropy of Ni crystals is four-fold symmetric, calculations are performed only in the first quadrant to increase calculation efficiency. The solid-phase region expands over time, indicating the progression of solidification. Additionally, growth in the horizontal direction is faster than in the 45-degree oblique direction, demonstrating anisotropy in crystal growth, which is a typical solidification structure.

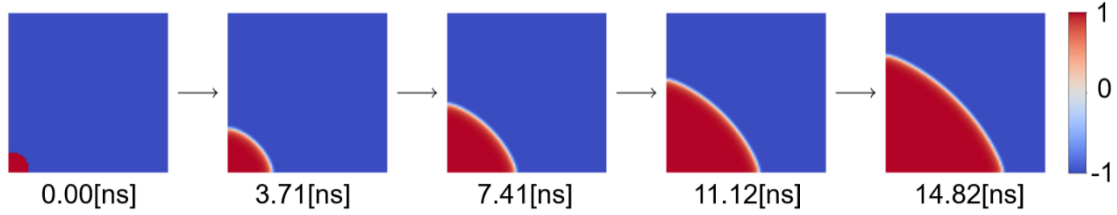

**Fig. 7.** Simulation results of Ni solidification using a quantitative phase-field model.

Table 4. Phase field parameter for Ni solidification.

| | | | |
|---|---|---|---|
| $T_m$ | Melting temperature | 1726 | [K] |
| $L$ | Latent heat | $2.311 \times 10^9$ | [J/m$^3$] |
| $c_p$ | Specific heat capacity | $5.313 \times 10^6$ | [J/m$^3$K] |
| $d_0$ | Capillary length | 0.5560 | [nm] |
| $\gamma_0$ | Interfacial energy | 0.326 | [J/m$^2$] |
| $\beta_0$ | Kinetic coefficient | 0.005 | [s/m] |
| $\varepsilon_c$ | Capillary anisotropy | 0.018 | [-] |
| $\varepsilon_k$ | Kinetic anisotropy | 0.130 | [-] |
| $u$ | Degree of supercooling | $-0.2$ | [-] |
| $dx$ | Grid size | 8.052 | [nm] |
| $dt$ | Time-step | 3.705 | [ps] |
| | Number of grids | $80 \times 80$ | [-] |

In this section, we conduct a twin experiment in which $\varepsilon_c$ and $\varepsilon_k$ are estimated by DA using 10 spatial distributions of the phase-field variable up to 5000 steps at 500 step intervals from above quantitative PFM simulation as the synthetic observation data. Note that these anisotropy parameters are challenging to measure using experimental approaches; therefore, most previous studies have relied on assumed or estimated values.

The state and observation vectors are defined using the vectored form of the phase-field variable, as follows:

$$x_{t|t} = \{\hat{\boldsymbol{\phi}} \quad \hat{\varepsilon}_c \quad \hat{\varepsilon}_k\}^T \tag{37}$$

$$y_t = \{\boldsymbol{\phi}\}^T \tag{38}$$

The procedure for converting the phase-field variable $\phi$ into its vector format $\boldsymbol{\phi}$ is illustrated in Figure 8 in the Supplementary Material. The DA parameters used for data assimilation are listed in Table 5, some of which are utilized in the following matrix form for estimation.

$$\boldsymbol{Q}_t = \text{diag}(Q_\phi, \cdots, Q_\phi, Q_{\varepsilon_c}, Q_{\varepsilon_k}) \tag{39}$$

$$\boldsymbol{R}_t = \text{diag}(R_\phi, \cdots, R_\phi) \tag{40}$$

$$\boldsymbol{H}_t = \begin{pmatrix} 1 & 0 & \cdots & 0 & 0 & \cdots & 0 \\ 0 & \ddots & \ddots & \vdots & \vdots & \ddots & \vdots \\ \vdots & \ddots & \ddots & 0 & \vdots & \ddots & \vdots \\ 0 & \cdots & 0 & 1 & 0 & \cdots & 0 \end{pmatrix} \tag{41}$$

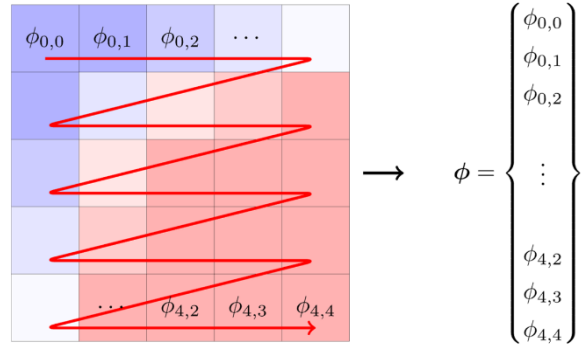

**Fig. 8.** Schematic diagram of the conversion from phase-field variable $\phi_{ij}$ into state vector $\boldsymbol{\phi}$.

**Table 5.** Data assimilation parameter for phase field Ni solidification.

| $N$ | Number of ensemble members | 10 |
|---|---|---|
| $k$ | Dimension of the state vector | 6402 |
| $l$ | Dimension of the observation vector | 6400 |
| $\Delta t_{\text{obs}}$ | Observation time interval | 1.85 [ns] |
| $Q_\phi$ | System noise covariance for $\phi$ | $1.0 \times 10^{-8}$ |
| $Q_{\varepsilon_c}$ | System noise covariance for $\varepsilon_c$ | $1.0 \times 10^{-12}$ |
| $Q_{\varepsilon_k}$ | System noise covariance for $\varepsilon_k$ | $1.0 \times 10^{-12}$ |
| $R_\phi$ | Observation noise covariance for $\phi$ | $1.0 \times 10^{-4}$ |
| $\boldsymbol{H}_t$ | Observation matrix | $\boldsymbol{I}_{6400,6402}$ |

Figure 9 shows results of the twin experiment by DA of the PFM. Estimated mean of the phase field variable from DA (Fig. 9(b)) is almost identical to synthetic observation data (Fig. 9(a)) in appearance, confirming that the DA reproduces the morphology of the crystals. Fig. 9(c) represents absolute difference between observation data and estimated phase field variable. It shows that regions with large differences between the estimated values and observational data are concentrated at the solid–liquid interface. This finding suggests that careful consideration of interface representation is crucial for accurately estimating interfacial properties using the PFM. Figure 9(d) and (e) represent the estimation results for $\varepsilon_c$ and $\varepsilon_k$. The results show that the estimated parameters converge to the true values used to generate the synthetic observation data, confirming that both the unknown parameters and the dynamic evolution of the crystal morphology were accurately estimated by this DA.

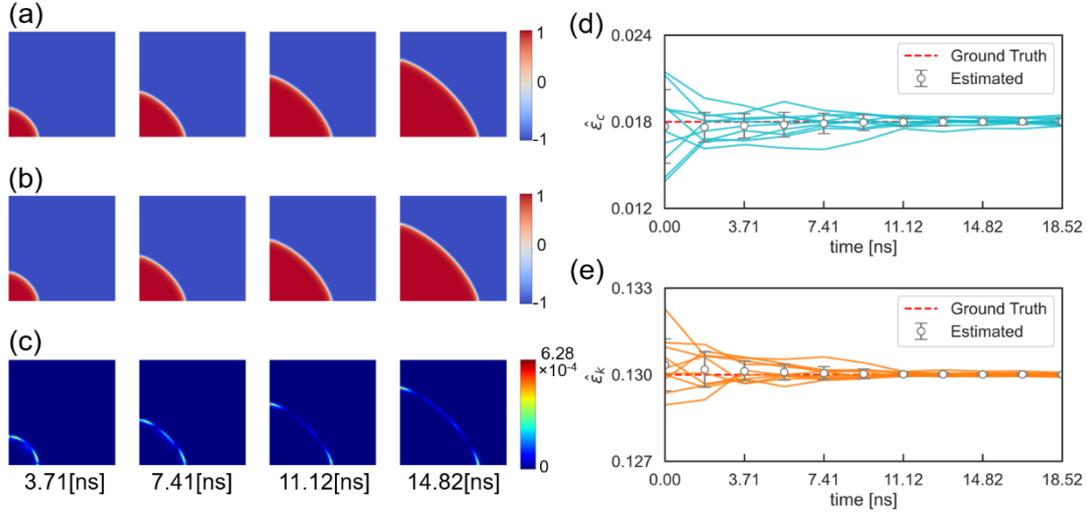

**Fig. 9.** Results of the twin experiment for estimating anisotropy parameters in Ni solidification using quantitative phase-field simulations. (a) Synthetic observation data. (b) Estimated mean of the phase-field variable obtained through data assimilation. (c) Absolute difference between the synthetic observation data and the estimated phase-field variable. (d, e) Temporal evolution of the estimated values of (d) capillary anisotropy $\hat{\varepsilon}_c$ and (e) kinetic anisotropy $\hat{\varepsilon}_k$. Dashed lines indicate the true values used to generate the synthetic observation data. Solid lines represent the estimated values from each ensemble member, while white dots denote the estimated mean.

**4.3 Data assimilation using experimental images: Volume-of-Fluid (VOF) Method**

We confirmed that DA was properly performed using douka in the above examples; however, all of these were twin experiments. The most exciting part of DA is the integration of practical experimental data with simulations. In this section, DA is performed using the images obtained in the experiment as observation data. Specifically, images taken by a high-speed camera of oxygen bubbles generated from an oxygen electrode during water electrolysis [37] were used as observation data, and DA with simulation using the Volume-of-Fluid (VOF) method was performed. We provide a brief background on this experimental system. The water electrolysis reaction has attracted attention as a clean energy storage technology for the development of a sustainable society. In this reaction, gas bubbles adhere to the electrode surface, hindering the reaction progress [38], which in turn reduces reaction efficiency. Therefore, analyzing the bubble detachment mechanism is crucial for improving reaction efficiency. In general, bubble detachment is explained by the balance between surface tension and buoyancy, with the detachment diameter typically considered to be around 1.6 mm [39]. However, experiments have reported the detachment of smaller bubbles that cannot be explained by this balance alone [40]. Elucidating this mechanism is expected to significantly enhance the efficiency of the reaction process. In order to bridge the gap between experiments and simulations, the process of bubble generation in water electrolysis was chosen as the target of DA. Details of the experimental setup for in-situ observation of $O_2$ bubble formation on a Ni electrode are provided in Appendix A.

The Volume-of-Fluid (VOF) method is a simulation technique used to analyze free surfaces in multiphase flows [41]. This method distinguishes fluid phases by introducing a volume fraction, denoted as $\alpha$ (VOF value). For example, in a two-phase flow consisting of air (gas phase) and water (liquid phase), the $\alpha$ value corresponds to the fluid phase as follows: $\alpha = 0$ represents air, while $\alpha = 1$ represents water, with the water surface exhibiting a continuous transition between 0 and 1. By computing the temporal evolution of $\alpha$, the fluid dynamics of multiphase flows can be simulated. The governing equations include the continuity equation and the Navier-Stokes equations (i.e., ordinary computation fluid dynamics (CFD)), with surface tension calculated using the continuum surface force model [42]. The time-dependent behavior of the fluid is then simulated by solving these equations.

In this section, we employ DA to investigate the oxygen gas supply rate, $U_z$, at the Ni electrode surface and the contact angle, $\theta$, at the electrode interface, both of which are considered crucial for understanding the bubble detachment mechanism. High-speed optical camera images of the oxygen bubble formation process [37] are used as observational data for DA. A large number of bubbles were generated throughout the experimental system; however, this study focuses on the growth process of a single bubble. The movie used for the actual observation data is provided in Supplementary Video S1. The images, captured at 1080 frames per second (fps), are processed by extracting bubble images every 100 frames and converting them to VOF values (i.e., continuous values between 0 and 1). The time interval between observational data points, $\Delta t_{obs}$, is approximately 92.6 ms. The image-to-VOF conversion is performed based on image contrast, followed by binarization and smoothing of the solid-liquid interface using a Gaussian filter. Specifically, a Gaussian filter with a radius of 1 pixel and a standard deviation of 1.0 is applied to the image pixel data. Figure 10 illustrates the original images and their corresponding VOF values. Among the 11 sampled VOF values, the one corresponding to the first image is used as the initial condition for the simulation, while the remaining ones serve as observational data for estimating the interfacial properties.

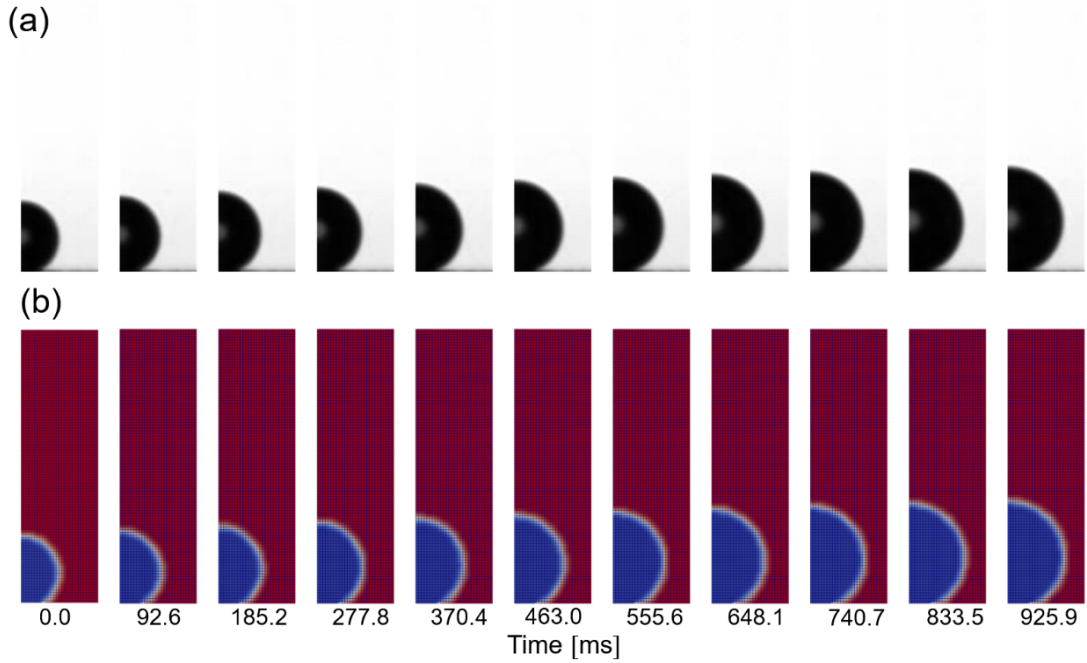

**Fig. 10.** (a) In-situ observation images of $O_2$ bubble formation process on Ni electrode [23]. (b) VOF values obtained by image transformation from in-site observation images.

The numerical simulations were conducted using OpenFOAM [43], an open-source CFD software. Notably, douka exhibits high versatility and can be integrated with existing software. To reduce computational demands, an axisymmetric computational domain is employed, centered on the forming bubble as depicted in Figure 11. The domain is discretized into a 100 × 100 × 1 grid. To model bubble growth, an $O_2$ gas inlet with radius $R$ is placed at the lower left of the domain. This inlet imposes a constant upward flow with velocity $U_z$ supplying $O_2$ gas into the domain. The remaining bottom surface is set with a boundary condition that imposes a contact angle $\theta$ for $O_2$. To prevent interface dissipation during bubble formation, the Geometric VOF method [44] is used for geometric interface reconstruction in the VOF calculations. The material properties and simulation parameters are detailed in Table 6.

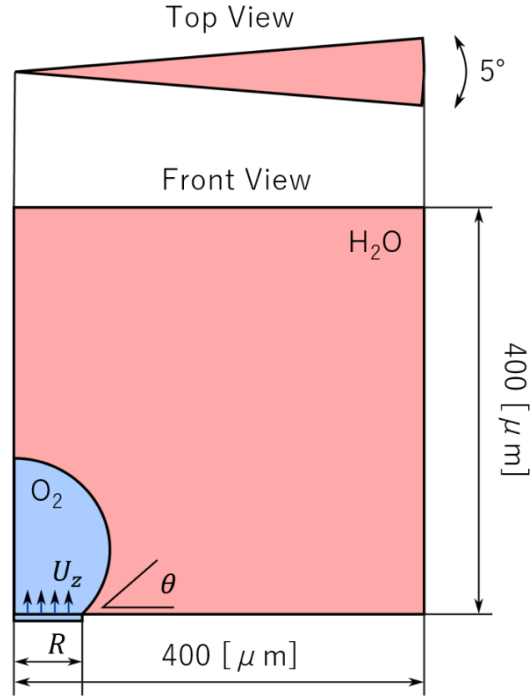

**Fig. 11.** Computational domain used for simulating the $O_2$ bubble formation process.

**Table 6.** Physical properties and calculation parameters used in the calculation of bubble formation

| | | | | |
|---|---|---|---|---|
| $\nu_1$ | Kinematic viscosity of $H_2O$ | $9.59 \times 10^{-7}$ | [m$^2$/s] | [44] |
| $\rho_1$ | Density of $H_2O$ | $1.0479 \times 10^3$ | [kg/m$^3$] | [44, 45] |
| $\nu_0$ | Kinematic viscosity of $O_2$ | $1.48 \times 10^{-5}$ | [m$^2$/s] | |
| $\rho_0$ | Density of $O_2$ | 1.429 | [kg/m$^3$] | |
| $\sigma$ | Surface tension | 0.0751 | [N/m] | [45] |
| $g$ | Gravitational acceleration | 9.81 | [m/s$^2$] | |
| $Co_{max}$ | Maximum Courant number | 0.5 | [-] | |
| $Co_{\alpha, max}$ | Maximum Courant number of $\alpha$ | 0.5 | [-] | |
| $M$ | KOH molar concentration | 0.9952 | [mol/L] | [44] |
| $R$ | Gas inlet radius | 32 | [μm] | |

The state and observation vectors for the DA are defined as follows:

$$x_{t|t} = \{\hat{\alpha} \quad \hat{p} \quad \hat{U} \quad \hat{U}_z \quad \hat{\theta}\}^T \tag{42}$$

$$y_t = \{\alpha\}^T \tag{43}$$

Here, $\boldsymbol{\alpha}$ reprensents the VOF value, $\boldsymbol{p}$ the pressure field and $\boldsymbol{U}$ the velocity vector field. The estimated oxygen gas inlet velocity $\widehat{U}_z$, and contact angle $\hat{\theta}$, are the parameters to be estimated. When using the observed VOF values as initial estimates, the pressure field at each domain is initialized with $p_0(\alpha - 1.0)$, where $p_0 = 1950.0$, to incorporate the bubble's internal pressure. The velocity field $\boldsymbol{U}$ is initialized to 0.0 throughout the domain. Table 7 details the parameters used in the DA of the bubble formation.

**Table 7.** Parameters used in data assimilation of bubble formation.

| | | |
|---|---|---|
| $N$ | Number of ensemble members | 10 |
| $k$ | Dimension of the state vector | 50002 |
| $l$ | Dimension of the observation vector | 10000 |
| $\Delta t_{\text{obs}}$ | Observation time interval | 92.6 [ms] |
| $Q$ | System noise covariance | 0.0 |
| $R_\alpha$ | Observation noise covariance for $\alpha$ | $1.0 \times 10^{-2}$ |
| $\boldsymbol{H}_t$ | Observation matrix | $\boldsymbol{I}_{l,k}$ |

Figure 12 presents the results of DA for bubble formation using experimental images. Notably, the estimated mean VOF value obtained from DA (Fig. 11(b)) closely aligns with the observational data from experimental images (Fig. 11(a)) in terms of appearance, confirming that DA successfully reproduces the morphology of bubble formation. The regions exhibiting significant discrepancies between the estimated values and the observational data are primarily concentrated at the bubble surface (Fig. 12(c)), demonstrating a similar trend to the twin experiment on crystal growth discussed in the previous section. Figure 12(d) and (e) represent the estimation results for $\widehat{U}_z$ and $\widehat{\theta}$ through DA. Both estimates exhibit a slight decrease in the initial stages but eventually converge to a constant value, demonstrating that DA can successfully estimate unknown parameters in CFD simulations as constant values based on experimental data. However, a discrepancy appears between the estimated contact angle and the one expected from visual inspection. Given that the contact angle $\theta$ represents a boundary condition in the CFD simulation (defining the angle at which the $O_2$ gas phase interacts with the boundary), the estimated value $\widehat{\theta}$ may not necessarily correspond to the actual contact angle of the bubble at the interface. Additionally, under the boundary conditions employed, the $O_2$ gas supply rate $U_z$ exerts an upward force on the entire bubble. As a result, the contact angle $\theta$ is expected to be influenced by $U_z$, particularly for small bubbles. Furthermore, Fig. 12(c) indicates that the largest absolute differences in VOF values are concentrated near the three-phase interface. Further improvements to the simulation model are necessary for more precise estimation of interfacial properties using experimental data, which is beyond the scope of this study. The most significant achievement of this research is the development of a general-purpose platform that enables DA using practical experimental images.

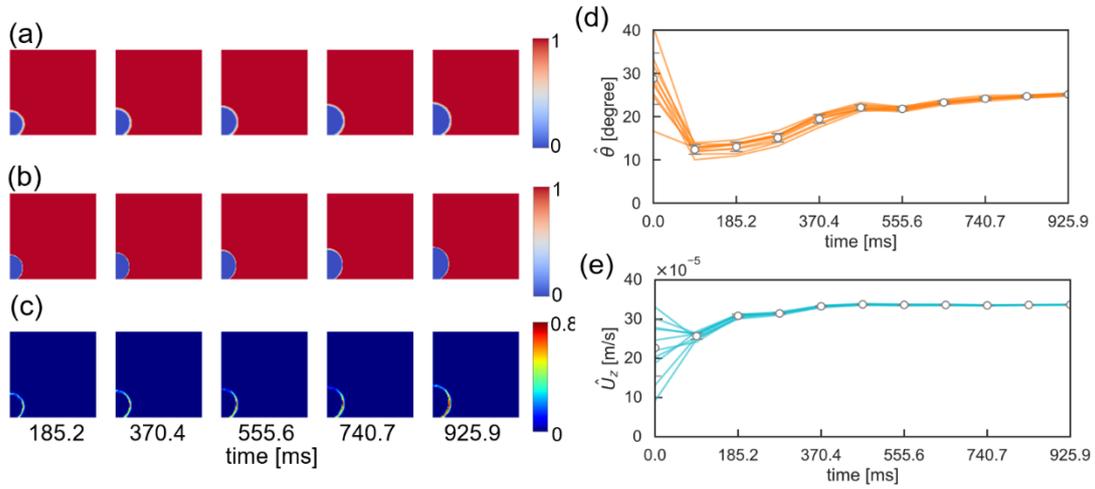

**Fig. 12.** Results of data assimilation for the $O_2$ bubble formation process on a Ni electrode using experimental observation data. (a) Observational data. (b) Estimated mean VOF value. (c) Absolute difference between observed and estimated VOF values. (d, e) Temporal evolution of the estimated values: (d) gas injection velocity ($\hat{U}_z$) and (e) contact angle ($\hat{\theta}$). Solid lines represent the estimated values for each ensemble member, while white dots indicate the estimated mean.

**4.4. Data assimilation with large ensemble members on a supercomputer Fugaku**

When performing sequential state estimation using nonlinear material models, it is essential to calculate the variance in the prediction step through ensemble approximation. In this section, we employ douka to conduct state estimation with large-scale ensemble members. Specifically, we focus on the estimation of interfacial properties using the PFM presented in Section 4.2, and we examine both the convergence of the estimation and the performance of the platform with large-scale ensemble members. All computations in this section were performed on the exascale supercomputer Fugaku at the RIKEN Center for Computational Science (R-CCS), Japan. Fugaku comprises 158,976 nodes, each equipped with a 48-core A64FX CPU and 32.0 GiB of memory. The target parameters for estimation are $\varepsilon_c$ and $\varepsilon_k$, and all computational conditions, except for the number of ensemble members, remain consistent with those described in Section 4.2. Twin experiments were performed to assess the prediction accuracy while varying the number of ensemble members from 2 to 8192, with one ensemble assigned to each node. As a benchmark, we measured the execution time for prediction. The runtime averaged across all ensemble members is regarded as the computation time for a single step.

Figures 13 and 14 present estimation results of capillary anisotropy $\hat{\varepsilon}_c$ and kinetic anisotropy $\hat{\varepsilon}_k$ for varying ensemble numbers, respectively. It is evident that as the ensemble size *N* increases, the deviation from the true value due to sampling errors in the initial estimation phase decreases. When the ensemble size is extremely small (e.g., $N = 2$), the variance of the estimated state converges rapidly, reducing the effectiveness of state correction by observational data. As a result, the true value may fall outside the estimated likelihood range. Figure 15 illustrates the execution time of the prediction step as a function of the ensemble size *N*. The execution time remains nearly constant and independent of *N*, demonstrating adherence to the ideal weak scaling law. In this twin experiment, the sampling error nearly converged at an ensemble size of approximately $N = 128$, suggesting that an ensemble size on the order of $10^2$ is sufficient. However, many cases require a larger number of ensemble members, such as those involving complex models with strongly nonlinear systems or DA using other methodology such as particle filter method. Therefore, the fact that the prediction step of douka constructed in this study achieves ideal weak scaling efficiency highlights its versatility for various applications.

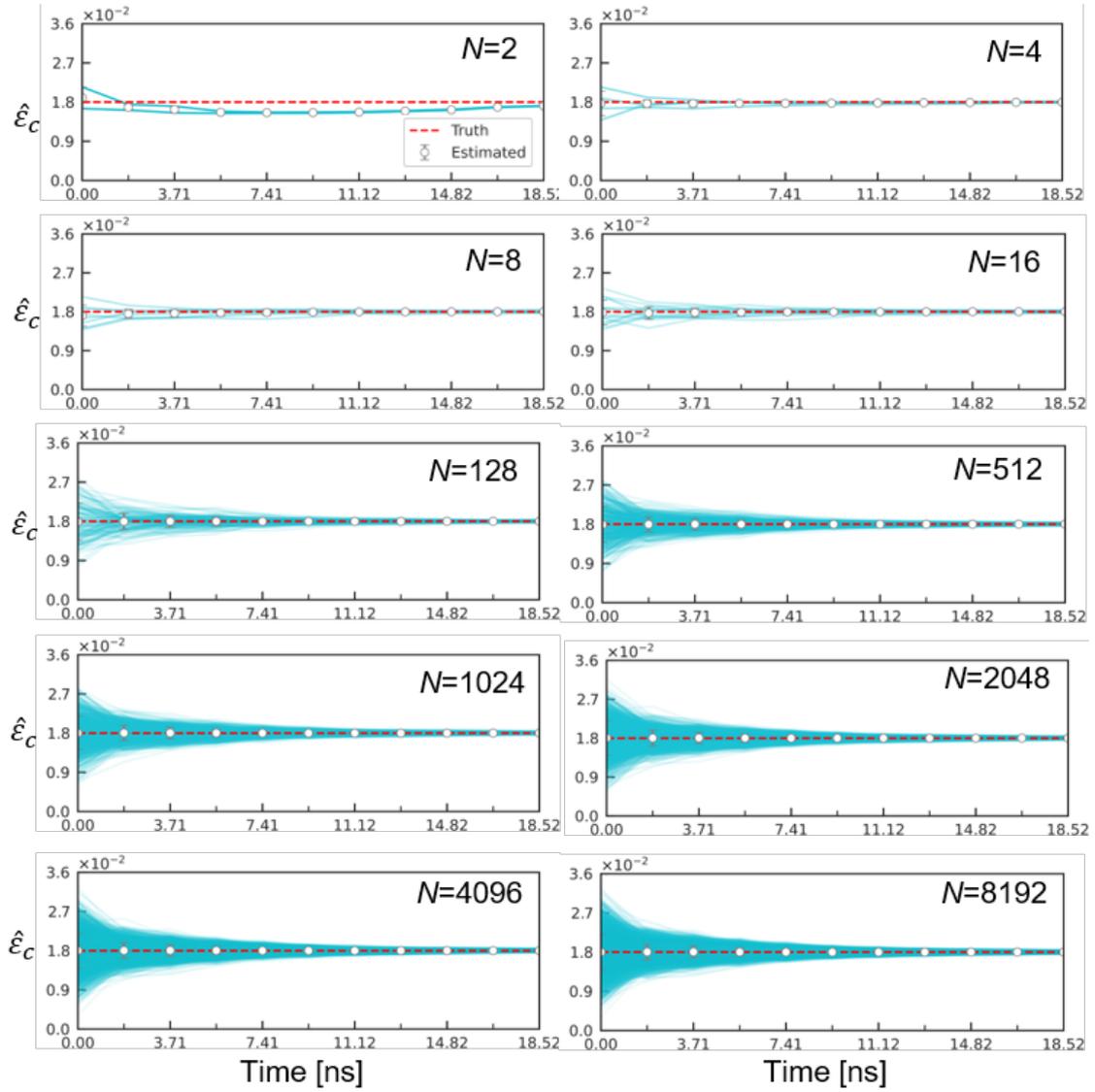

**Fig. 13.** Estimation result of capillary anisotropy $\hat{\varepsilon}_c$ for varying ensemble numbers $N$ by the large-scale data assimilation performed on the supercomputer, Fugaku. Dashed lines indicate the true values used to generate the synthetic observation data. Solid lines represent the estimated values from each ensemble member, while white dots denote the estimated mean.

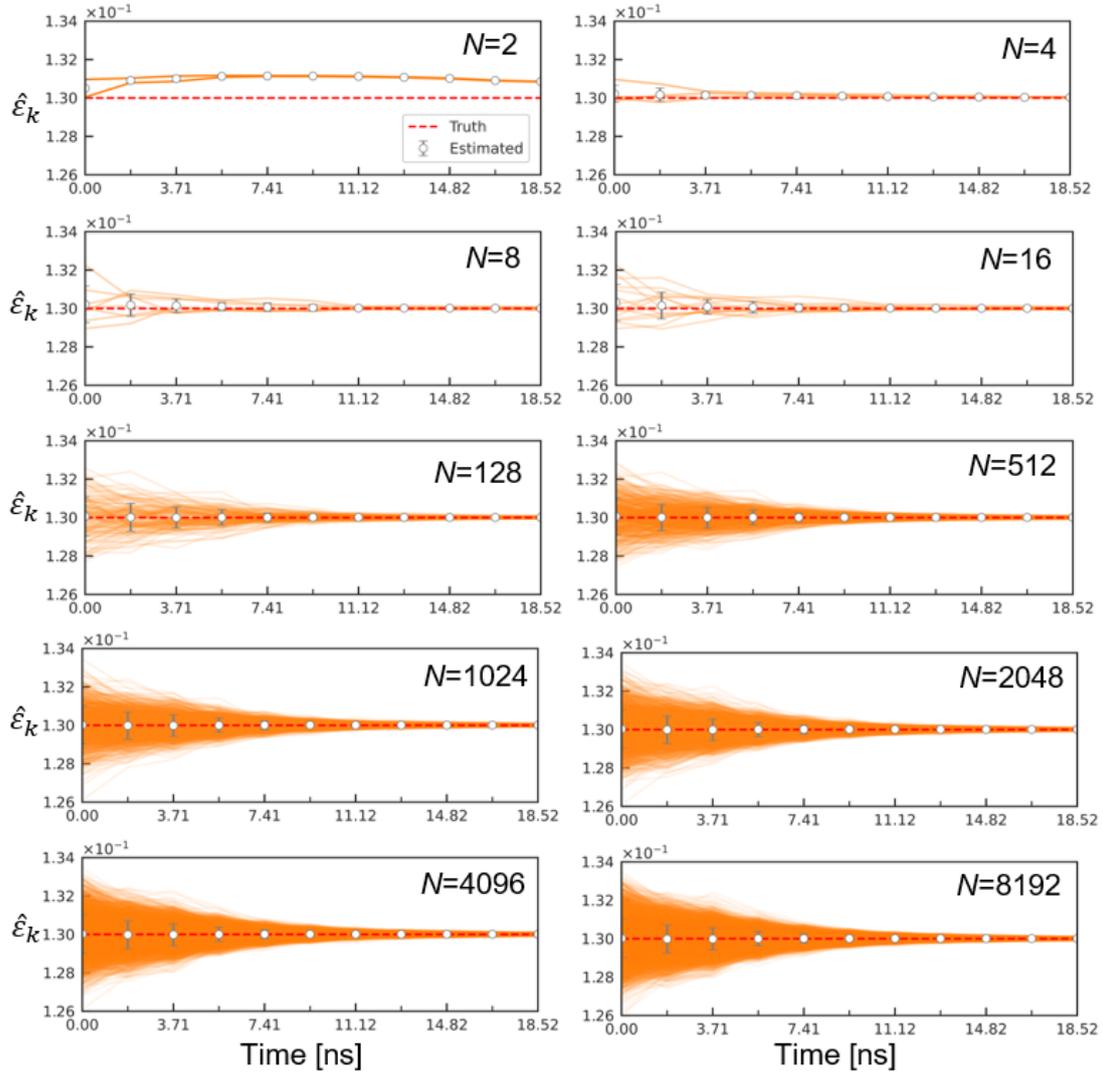

**Fig. 14.** Estimation result of kinetic anisotropy $\hat{\varepsilon}_k$ for varying ensemble numbers $N$ by the large-scale data assimilation performed on the supercomputer, Fugaku. Dashed lines indicate the true values used to generate the synthetic observation data. Solid lines represent the estimated values from each ensemble member, while white dots denote the estimated mean.

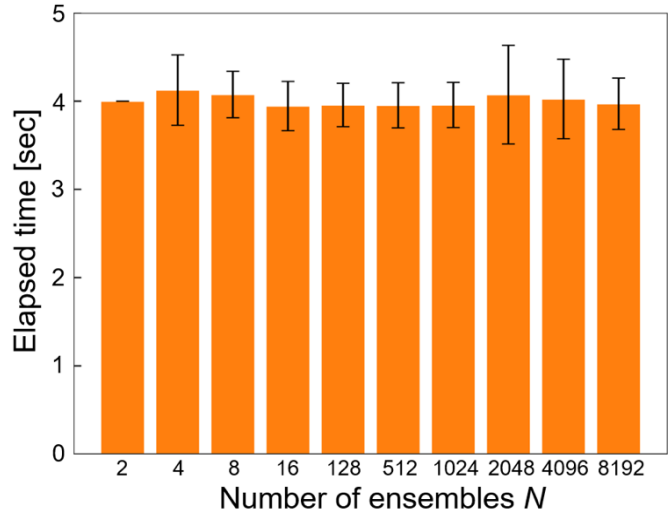

**Fig. 15.** The prediction command execution time as a function of ensemble numbers *N* by the large-scale data assimilation performed on the supercomputer, Fugaku.

## 5. Conclusions

In this study, a large-scale, general-purpose DA platform for materials modeling, named douka, was developed. Several applications of douka to nonlinear materials models are presented, demonstrating the effectiveness of DA. The platform's applicability to models with state vectors exceeding 1,000 dimensions is also shown, highlighting its capability to estimate physical properties that cannot be directly obtained from observed data. Notably, DA was successfully performed using experimental images of oxygen evolution at a water electrolysis electrode, enabling the estimation of injection velocity of oxygen gas and contact angle of the bubble.

Furthermore, this platform was applied to large-scale ensemble DA using the supercomputer Fugaku, achieving state estimation with up to 8,192 ensemble members. The results demonstrate that even as the number of ensemble members increases, the runtime for prediction step follows the weak scaling law, leading to a significant speedup. This finding suggests that the proposed approach effectively mitigates the rise in computational cost associated with ensemble size, even when applied to DA for material models requiring time-intensive analyses. The outcomes of this study are expected to contribute to the advancement of data-driven materials science. Future developments will focus on integrating additional DA methodologies, such as the particle filter and Ensemble Kalman smoothers. Additionally, enhancing the platform's scalability through the parallelization of filtering operations will be a key objective.

**Data availability**

The codes developed in this study are available on GitHub (https://github.com/ MaterialsModellingLab/douka) (v0.0.0 is released at https://doi.org/10.5281/zenodo. 15221070 ), including those for generating synthetic observation data for example applications.

**CRediT authorship contribution statement**

**Aoi Watanabe:** Data curation, Formal analysis, Investigation, Methodology, Software, Validation, Visualization, Writing – original draft, Writing – review & editing. **Sato Ryuhei:** Formal analysis, Methodology, Validation, Visualization, Writing – original draft, Writing – review & editing. **Ikuya Kinefuchi:** Conceptualization, Data curation, Funding acquisition, Resources, Validation, Visualization, Writing – original draft, Writing – review & editing. **Yasushi Shibuta:** Conceptualization, Formal analysis, Funding acquisition, Methodology, Project administration, Resources Supervision, Validation, Writing – original draft, Writing – review & editing.

**Declaration of competing interest**

The authors declare that they have no known competing financial interests or personal relationships that could have appeared to influence the work reported in this paper.

**Acknowledgments**

The authors would like to thank Yusuke Suzuki, Kohei Nakano and Kohei Sato at Department of Mechanical Engineering, the University of Tokyo for the assistance in preparing experimental data of oxygen evolution reaction used in Section 4.3. This work was supported by MEXT Program: Data Creation and Utilization-Type Material Research and Development Project Grant Number JPMXP1122712807. Part of this work was supported by MEXT as "Program for Promoting Researches on the Supercomputer Fugaku"(Data-Driven Research Methods Development and Materials Innovation Led by Computational Materials Science, JPMXP1020230327) and used computational resources of supercomputer Fugaku provided by the RIKEN Center for Computational Science (Project ID: hp230212, hp240223 and hp250229).

**Appendix A. Experimental setup for in-situ observation of $O_2$ bubble formation on a Ni electrode**

Electrolysis was performed using a nickel film as the anode and a nickel wire as the cathode. The nickel film electrode was patterned using the following procedure. First, a photoresist was spin-coated onto a glass substrate. The photoresist was then patterned using the photolithography technique. Subsequently, a thin nickel film with a thickness of 500 nm was sputtered onto the substrate with the patterned photoresist. Finally, the lift-off process was employed to obtain the patterned electrode with a width of 50 μm. A 1 mol/L KOH aqueous solution was used as the electrolyte. The temperature of the solution, measured with a K-type thermocouple, was maintained at 23.4 ± 0.2 °C. A constant voltage was applied to the electrodes using a DC power supply. The applied voltage was adjusted so that the current density of the anode became a target value between 0.1–3 A/cm². Oxygen bubbles generated on the anode were recorded using a high-speed camera (Photoron FASTCAM5) and a macro lens (LAOWA 25mm F2.8 2.5–5X ULTRA MACRO) at the frame rate of 1080 fps. To ensure sufficient brightness for high-speed imaging, a 3W white LED was used for illumination. Figure A1 shows the experimental setup.

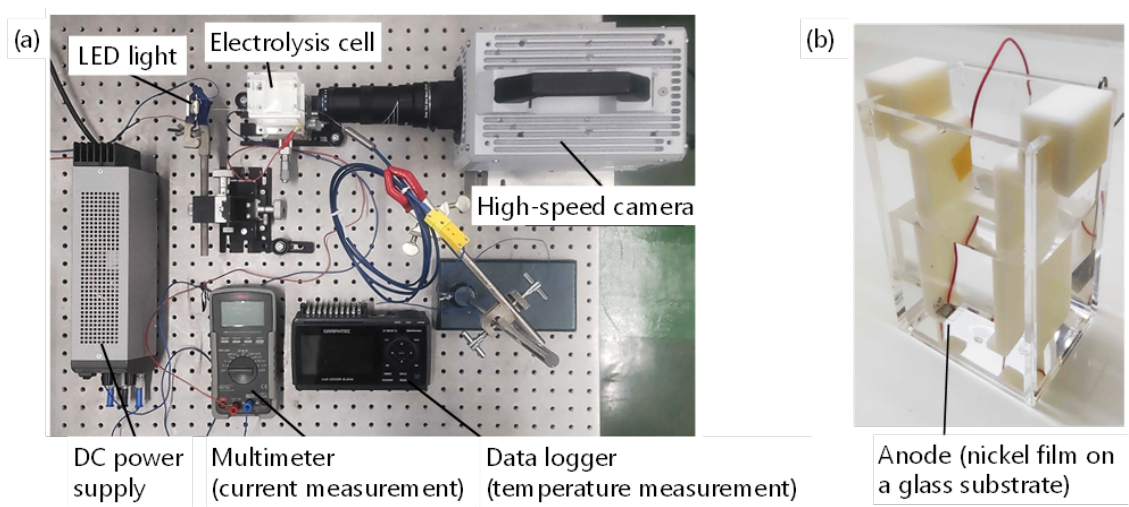

**Fig. A1.** Experimental setup. (a) Overview of the entire setup. (b) Close-up view of the electrolysis cell.